\documentstyle[aps,eqsecnum,preprint]{revtex}
\begin{document}
\draft
\preprint{WIS -- 95/52/Oct -- PH}
\title{Microscopic derivation of rate equations for quantum transport}
\author{S. A. Gurvitz and Ya. S. Prager}
\address{Department of Particle Physics, Weizmann Institute of
         Science, Rehovot 76100, Israel}
\date{\today}
\maketitle
\begin{abstract}
It is shown that under certain conditions the resonant transport
in mesoscopic systems can be described by modified (quantum) rate equations,
which resemble the optical Bloch  equations with some additional 
terms. Detailed microscopic derivation from the many-body Schr\"odinger
equation is presented. Special attention is paid to the Coulomb
blockade and quantum coherence effects in coupled quantum dot systems.
The distinction between classical and quantum descriptions of 
resonant transport is clearly manifested in the modified rate equations.
\end{abstract}

\section{Introduction}
Over the last decade a great interest has been paid to    
artificially fabricated nanostructures containing discrete number of 
quantum states. The discreteness of quantum states 
manifests itself in peculiar transport properties of these systems 
as, for instance, in the Coulomb blockade oscillations\cite{likh}.
Actually, the study has been mostly concentrated on the quantum transport 
through single devices (quantum dots). In fact, more interesting 
quantum mechanical effects can be found in {\em coupled} 
nanostructures devices,  
where the quantum interference may strongly influence the resonance current.  
The impressive progress in microfabrication technology now allows to  
extend the experimental investigation to these systems too. 
For instance, the transport properties of coupled dots 
are presently under intensive study\cite{vdr,wbm}. 

For description of quantum transport through single quantum dot
(quantum well) the ``classical'' rate equations are usually 
used\cite{glaz,ak,been}. They can be derived either by
using nonequilibrium Green's functions technique\cite{Davies4603}, or 
directly from the Schr\"odinger equation\cite{glp}. 
The situation is different for coupled wells with aligned levels.
The quantum transport through these devices goes on via quantum 
superposition between the states in adjacent wells.  
It is thus quite obvious that 
non-diagonal density matrix elements would appear in the equations 
of motion. These terms have no classical
counterparts, and therefore the classical rate equations have to be modified. 
A plausible modification of master equations for some particular 
cases of the resonance tunneling through double-dot structures
has been proposed by Nazarov\cite{naz}. A more general case,
though without account of Coulomb interaction, has been considered 
in\cite{glp1}, where modified rate equations have been proposed 
by using an analogy to the optical Bloch equations\cite{bloch}.
However, no microscopic derivation of the modified rate equations 
has been presented yet.

In this paper we derive the rate equations for a general case of 
resonant transport through mesoscopic systems, starting 
with the many-body Schr\"odinger equation, with special attention being 
paid to the Coulomb blockade and coherent quantum mechanical effects.
Our main goals are, first, to substantiate and generalize the
previously suggested rate equations 
and second, to determine the region of validity of the rate 
equations for description of quantum transport in general. 
Also, we believe that the microscopic derivation of the rate equations 
will provide a better understanding of the correspondence between   
quantum and classical description of carrier transport in mesoscopic systems. 

The plan of the paper is the following. In Sect.~2 we give a detailed 
derivation of the transport rate equations through a single quantum well (dot).
In order to present our method most lucidly, we neglect 
in this section the Coulomb interaction and spin effects. These effects 
are considered in Sect.~3. In Sect.~4 we derive the modified rate 
equations for coupled well structures, taking into account the Coulomb 
and spin effects. An example of coherent resonant transport with inelastic 
transitions is studied in Sect.~5. The derivation of rate equations performed 
in this case allows to establish their correct form 
valid in general case of quantum transport. The general case and
an example of coherent resonant transport with inelastic 
transitions in the presence of strong Coulomb blockade
are presented in Sect.~6. The last section is a summary. 

\section{Single-well structure} 
Let us consider a mesoscopic ``device'' consisting of a quantum
well (dot), coupled to two separate electron reservoirs. 
The density of states in the reservoirs is very high (continuum). 
The dot, however, contains only isolated levels. 
We first demonstrate how to achieve the reduction of many-body 
Schroedinger equation to the rate equation in the simplest example, Fig. 1,
with only one level, $E_1$, inside the dot. 
We also ignore the Coulomb electron-electron interaction 
inside the well and the spin degrees of freedom. 
Hence, only one 
electron may occupy the well. With the stand simplifications, the tunneling
Hamiltonian of the entire
system in the occupation number representation is
\begin{eqnarray}
{\cal H} =\sum_l E_{l}a^{\dagger}_{l}a_{l} +
E_1 a_1^{\dagger}a_{1} +\sum_r E_{r}a^{\dagger}_{r}a_{r} 
+ \sum_l \Omega_{l}(a^{\dagger}_{l}a_1 +a^{\dagger}_{1}a_{l})
+ \sum_r \Omega_{r}(a^{\dagger}_{r}a_1 +a^{\dagger}_{1}a_{r})\;.
\label{Ham}
\end{eqnarray}
Here the subscripts $l$ and
$r$ enumerate correspondingly the (very dense) levels in the
left (emitter) and right (collector) reservoirs. 
For simplicity,  we restrict ourselves to the zero temperature case.
All the levels in the emitter and the 
collector are initially filled with electrons up to 
the Fermi energy $E_F^L$ and $E_F^R$, respectively. 
This situation will be treated as the ``vacuum'' state $|0\rangle$.
 
This vacuum state is unstable; the Hamiltonian
Eq.~(\ref{Ham}) requires it to decay exponentially to a continuum state
having the form $a_{1}^{\dagger}a_{l}|0\rangle$ with an electron in
the level $E_1$ and a hole in the emitter continuum. These continuum
states are also unstable and decay to states
$a_{r}^{\dagger}a_{l} |0\rangle$
having a particle in the
collector continuum as well as a hole in the emitter continuum, and
no electron in the level $E_1$. The latter, in turn, are decaying into
the states $a_{1}^{\dagger}a_{r}^{\dagger}a_{l}a_{l'} |0\rangle$
and so on. The evolution of the whole system is described by the
many-particle wave function, which is represented as
\begin{eqnarray}
|\Psi (t)\rangle = \left [ b_0(t) + \sum_l b_{1l}(t)
           a_{1}^{\dagger}a_{l} + \sum_{l,r} b_{lr}(t)
           a_{r}^{\dagger}a_{l}
          + \sum_{l<l',r} b_{1ll'r}(t)
           a_{1}^{\dagger}a_{r}^{\dagger}a_{l}a_{l'}
           + \ldots \right ] |0\rangle, 
\label{wf}
\end{eqnarray}
where $b(t)$ are the time-dependent probability amplitudes to
find the system in the corresponding states described above with the initial 
condition $b_0(0)=1$, and all the other $b(0)$'s being zeros.
Substituting Eq.~(\ref{wf}) into in the
Schr\"odinger equation $i|\dot\Psi (t)\rangle ={\cal H}|\Psi (t)\rangle$,
results in an infinite set of coupled linear differential equations for the
amplitudes $b(t)$. Applying the Laplace transform 
\begin{equation}
\tilde{b}(E)=\int_0^{\infty}e^{iEt}b(t)dt
\label{lap}
\end{equation}
and taking account of the initial conditions, we transform 
the linear differential 
equations for $b(t)$ into an infinite set of algebraic equations 
for the amplitudes $\tilde b(E)$, 
\begin{mathletters}
\label{ineq}
\begin{eqnarray}
& &E \tilde{b}_{0}(E) - \sum_l \Omega_{l}\tilde{b}_{1l}(E)=i
\label{ineq1}\\
&(&E + E_{l} - E_1) \tilde{b}_{1l}(E) - \Omega_{l}
      \tilde{b}_{0}(E) - \sum_r \Omega_{r}\tilde{b}_{lr}(E)=0
\label{ineq2}\\
&(&E + E_{l} - E_{r}) \tilde{b}_{lr}(E) -
      \Omega_{r} \tilde{b}_{1l}(E) - 
      \sum_{l'} \Omega_{l'}\tilde{b}_{1ll'r}(E)=0
\label{ineq3}\\
&(&E + E_{l} + E_{l'} - E_1 - E_{r}) \tilde{b}_{1ll'r}(E)-
\Omega_{l'} \tilde{b}_{lr}(E)+
\Omega_{l} \tilde{b}_{l'r}(E)-
\sum_{r'} \Omega_{r'}\tilde{b}_{ll'rr'}(E)=0
\label{ineq4}\\
& &\cdots\cdots\cdots\cdots\cdots 
\nonumber
\end{eqnarray}
\end{mathletters}

Eqs. (\ref{ineq}) can be substantially simplified. Let us replace  
the amplitude $\tilde b$ in the term $\sum\Omega\tilde b$ 
of each of the equations  by 
its expression obtained from the subsequent equation. For example,   
substitute $\tilde{b}_{1l}(E)$ from Eq.~(\ref{ineq2}) into Eq.~(\ref{ineq1}).  
We obtain
\begin{equation}
\left [ E - \sum_l \frac{\Omega^2_{l}}{E + E_{l} - E_1}
    \right ] \tilde{b}_{0}(E) - \sum_{l,r}
    \frac{\Omega_{l}\Omega_{r}}{E + E_{l} - E_1}
    \tilde{b}_{lr}(E)=i.
\label{exam}
\end{equation}  
Since the states in the reservoirs are very dense (continuum), 
one can replace the sums over $l$ and $r$ by integrals, for instance  
$\sum_{l}\;\rightarrow\;\int \rho_{L}(E_{l})\,dE_{l}\:$,
where $\rho_{L}(E_{l})$ is the density of states in the emitter. 
Then the first sum in Eq.~(\ref{exam}) becomes an
integral which can be split into a sum of the singular and principal value 
parts. The singular part
yields $\;\,-i\Theta (E_F^L+E-E_1)\,\Gamma_L/2$, where $\Gamma_L = 2\pi
\rho_L(E_1)|\Omega_L(E_1)|^2$ is the level $E_1$ partial width 
due to coupling to the emitter. Let us assume that  
$E_F^L\gg E_1\gg E_F^R$, i.e. the bias is large and 
the energy level is deeply inside the band.  
In this case the integration over $E_{l(r)}$-variables can be extended 
to $\pm\infty$. As a result, the theta-function can be replaced by one, and   
the principal part is merely included into
redefinition of the energy $E_1$. Also, 
the second sum (integral) in Eq.~(\ref{exam}) proves to be
negligibly small. Indeed, let us replace   
$\tilde{b}_{lr}\to\tilde{b}(E_l,E_r,E)$, and assume weak energy
dependence of $\Omega$ on $E_{l(r)}$. Then 
one finds from Eqs. (\ref{ineq}) that the poles of the integrand in the 
$E_l(E_r)$-variable are on one side of the integration contour, 
and therefore this term vanishes. 
 
Applying analogous considerations to the other equations of the
system (\ref{ineq}), we finally arrive to the following set of equations: 
\begin{mathletters}
\label{fineq}
\begin{eqnarray}
&& (E + i\Gamma_L/2) \tilde{b}_{0}(E)=i
\label{fineq1}\\
&& (E + E_{l} - E_1 + i\Gamma_R/2) \tilde{b}_{1l}(E)
      - \Omega_{l} \tilde{b}_{0}(E)=0
\label{fineq2}\\ 
&& (E + E_{l} - E_{r} + i\Gamma_L/2) \tilde{b}_{lr}(E) -
      \Omega_{r} \tilde{b}_{1l}(E)=0
\label{fineq3}\\
&& (E + E_{l} + E_{l'} - E_1 - E_{r} + i\Gamma_R/2) 
       \tilde{b}_{1ll'r}(E) - 
       \Omega_{l'} \tilde{b}_{lr}(E) +\Omega_{l} \tilde{b}_{l'r}(E)=0
\label{fineq4}\\
& &\cdots\cdots\cdots\cdots\cdots 
\nonumber
\end{eqnarray}
\end{mathletters}
where $\Gamma_R = 2\pi\rho_R(E_1)|\Omega_R(E_1)|^2$ is 
the level $E_1$ partial width due to coupling to the collector.   

Now we introduce the density matrix of the ``device''. 
The Fock space of the quantum well consists of only two possible states, 
namely: $|a\rangle$ -- the level $E_1$ is empty,
and $|b\rangle$ -- the level $E_1$ is occupied. In this basis,
the diagonal elements of the density matrix of the ``device'',
$\sigma_{aa}$ and $\sigma_{bb}$, give the probabilities of the
resonant level being empty or occupied, respectively. In our 
notation, these probabilities are represented as follows:
\begin{mathletters}
\label{sigmas}
\begin{eqnarray}
\sigma_{aa} &=& |b_{0}(t)|^2 + \sum_{l,r} |b_{lr}(t)|^2
            + \sum_{l<l',r<r'} |b_{ll'rr'}(t)|^2 + \ldots
\nonumber\\
          &\equiv& \sigma_{aa}^{(0)} + \sigma_{aa}^{(1)}
            + \sigma_{aa}^{(2)} + \ldots\;,
\label{sigmaa}\\
\sigma_{bb} &=& \sum_l |b_{1l}(t)|^2 +
             \sum_{l<l',r} |b_{1ll'r}(t)|^2 
             + \sum_{l<l'<l'',r<r'} |b_{1ll'l''rr'}(t)|^2 + \ldots
\nonumber\\
             &\equiv& \sigma_{bb}^{(0)} + \sigma_{bb}^{(1)}
             + \sigma_{bb}^{(2)} + \ldots,
\label{sigmbb}
\end{eqnarray}
\end{mathletters}
where the index $n$ in $\sigma^{(n)}$ denotes the number of electrons 
in the collector. The current  $I(t)$ flowing through the system is  
$I(t)=e\dot N_R(t)$, where $N_R(t)$ is the number of 
electrons accumulated in the collector, i.e. 
\begin{equation}
N_R(t) = \sum_{n} n\left [\sigma_{aa}^{(n)}(t)+\sigma_{bb}^{(n)}(t)\right ]
\label{char}
\end{equation}

The density submatrix elements are directly related 
to the amplitudes $\tilde b(E)$ through the inverse Laplace transform 
\begin{equation}
\sigma^{(n)}(t)=
\sum_{l\ldots , r\ldots}\int \frac{dEdE'}{4\pi^2}\tilde b_{l\cdots r\cdots}(E)
\tilde b^*_{l\cdots r\cdots}(E')e^{i(E'-E)t}
\label{invlap}
\end{equation}
By means of this equation one can transform  Eqs. (\ref{fineq}) for 
the amplitudes $b(E)$  
into differential equations directly for the probabilities $\sigma^{(n)}(t)$. 
Consider, for instance, the term $\sigma_{bb}^{(0)}(t)=\sum_l |b_{1l}(t)|^2$,
Eq.~(\ref{sigmbb}). Multiplying Eq.~(\ref{fineq2}) by $\tilde{b}^*_{1l}(E')$ 
and then subtracting the complex conjugated equation with the interchange 
$E\leftrightarrow E'$ we obtain 
\begin{eqnarray}
\int\frac{dEdE'}{4\pi^2}(E'-E
-i\Gamma_R)\sum_l&&\tilde b_{1l}(E)\tilde b^*_{1l}(E')e^{i(E'-E)t}\nonumber\\
-&&\int\frac{dEdE'}{4\pi^2}2{\mbox {Im}}\sum_l\Omega_l
\tilde b_{1l}(E)\tilde b^*_0(E')e^{i(E'-E)t}=0
\label{a1}
\end{eqnarray}
One can easily deduce from Eq.~(\ref{invlap}) that the first integral in 
Eq.~(\ref{a1}) equals to $-i[\dot\sigma_{bb}^{(0)}(t)+
\Gamma_R\sigma_{bb}^{(0)}(t)]$. Next, substituting  
\begin{equation}
\tilde{b}_{1l}(E)=\frac{\Omega_{l} \tilde{b}_{0}(E)}
{E + E_{l} - E_1 + i\Gamma_R/2}
\label{a2}
\end{equation}
from Eq.~(\ref{fineq2}) into the second integral of Eq.~(\ref{a1}),
and replacing $\sum_l\to\int\rho_L(E_l)dE_l$ one can can perform  
the $E_l$-integration in the integral, thus obtaining 
$i\Gamma_L\sigma_{aa}^{(0)}(t)$.  
Ultimately, Eq.~(\ref{a1}) reads 
 $\dot{\sigma}^{(0)}_{bb}(t)=\Gamma_L \sigma^{(0)}_{aa}(t)
- \Gamma_R \sigma_{bb}^{(0)}(t)$. We can go on with this algebra for  
all the other amplitudes $\tilde b(t)$. For instance,
the above procedure applied to Eq.~(\ref{fineq4}) converts it into 
a differential equation for the density-matrix element
$\sigma_{bb}^{(1)}$, Eq.~(\ref{sigmbb}). The only difference with the
previous example is an appearance of the ``cross terms'', like 
$\sum\Omega_{l} \tilde{b}_{l'r}(E)\Omega_{l'} \tilde{b}^*_{lr}(E')$.
Yet, these terms vanish after the integration over $E_{l(r)}$, 
just as the second term in Eq.~(\ref{exam})).  
The rest of the algebra remains the same, so one obtains
$\dot{\sigma}^{(1)}_{bb}(t)=\Gamma_L \sigma^{(1)}_{aa}(t)
- \Gamma_R \sigma_{bb}^{(1)}(t)$. 
Finally we arrive to the following infinite system of the 
chain equations for the diagonal elements, 
$\sigma_{aa}^{(n)}$ and $\sigma_{bb}^{(n)}$, of the density matrix, 
\begin{mathletters}
\label{aandb}
\begin{eqnarray}
& &\dot{\sigma}^{(0)}_{aa}(t) = - \Gamma_L \sigma^{(0)}_{aa}(t)\;,
\label{anought}\\
& &\dot{\sigma}^{(0)}_{bb}(t) = \Gamma_L \sigma^{(0)}_{aa}(t)
                               - \Gamma_R \sigma_{bb}^{(0)}(t)\;,
\label{bnought}\\
& &\dot{\sigma}^{(1)}_{aa}(t) = - \Gamma_L \sigma^{(1)}_{aa}(t)
                               + \Gamma_R \sigma_{bb}^{(0)}(t)\;,
\label{aone}\\  
& &\dot{\sigma}^{(1)}_{bb}(t) = \Gamma_L \sigma^{(1)}_{aa}(t)
                               - \Gamma_R \sigma_{bb}^{(1)}(t)\;,
\label{bone}\\
& &\cdots\cdots\cdots\cdots\cdots 
\nonumber
\end{eqnarray}
\end{mathletters}
Summing up these equations, one easily obtains 
differential equations for the 
total probabilities $\sigma_{aa}=\sum_n \sigma_{aa}^{(n)}$ and 
 $\sigma_{bb}=\sum_n \sigma_{bb}^{(n)}$:
\begin{mathletters}
\label{a4}
\begin{eqnarray}
\dot\sigma_{aa} & = & -\Gamma_L\sigma_{aa}+\Gamma_R\sigma_{bb}\;,
\label{a4a}\\
\dot\sigma_{bb} & = & \Gamma_L\sigma_{aa}-\Gamma_R\sigma_{bb}\;,
\label{a4b}
\end{eqnarray}
\end{mathletters}  
which should be supplemented with the initial conditions
\begin{equation}
\sigma_{aa}(0)=1, \;\;\;\;\sigma_{bb}(0)=0.
\label{init}
\end{equation}

Using Eqs. (\ref{char}), (\ref{aandb}) we obtain  
the total current
\begin{equation}
I(t)=e\dot N_R(t)= e\Gamma_R [\sigma_{bb}^{(0)}(t) + \sigma_{bb}^{(1)}(t) 
       + \sigma_{bb}^{(2)}(t) + \ldots ] = e\Gamma_R \sigma_{bb}(t).
\label{curr}
\end{equation}
Thus the current $I(t)$ is directly proportional to 
the charge density in the well. Solving Eqs.~(\ref{a4}) 
and substituting $\sigma_{bb}(t)$ into Eq.~(\ref{curr}), we obtain (for
$t\to\infty$) the standard formula for the dc resonant
current,
\begin{equation}
I/e=\frac{\Gamma_L\Gamma_R}{\Gamma_L+\Gamma_R}\;.
\label{a5}
\end{equation}
Notice that whereas the time-behavior of the current $I(t)$ depends on the 
initial condition,   
the stationary current $I=I(t\to\infty)$, Eq.~(\ref{a5}), does not. 

Equations (\ref{a4}), derived from the many-body Schr\"odinger equation,
coincide with the classical rate equations in the sequential picture for 
the resonant tunneling, obtained using 
nonequilibrium quantum statistical mechanics technique\cite{Davies4603}.   
In contrast, our approach starts directly from the many-body 
Schr\"odinger equation and will be straightforwardly extended to
more complicated situations. 
Note, however, that the method can be applied only  
when the resonance energy is inside the band, and 
$\Gamma_{L,R}\ll E_F^L-E_F^R$. If the resonance is near the 
edge of the band, but the width of the
resonance is much smaller than the band width, our method still 
can be applied, but only to the stationary case ($t\to\infty$). Yet,  
the time-dependent Scr\"odinger equation cannot be reduced to
the rate equations (\ref{a4}), and therefore this case is not 
a subject of this paper.

\section{Coulomb blockade}
Now we extend the approach of Sect.~2 to include the 
effects of Coulomb interaction. Consider again the 
quantum well in Fig. 1, taking into 
account the spin degrees of freedom ($s$). In this case 
the tunneling Hamiltonian (\ref{Ham}) becomes
\begin{eqnarray}
{\cal H} &=& \sum_{l,s} E_{l}a^{\dagger}_{ls}a_{ls} +
              \sum_{s}E_1 a^{\dagger}_{1s}a_{1s} +
              \sum_{r,s} E_{r}a^{\dagger}_{rs}a_{rs} 
\nonumber\\
          & & + \sum_{l,s} \Omega_{l}(a^{\dagger}_{ls}a_{1s} +
              a^{\dagger}_{1s}a_{ls})
              + \sum_{r,s} \Omega_{r}(a^{\dagger}_{rs}a_{1s} +
              a^{\dagger}_{1s}a_{rs})
             +Ua^{\dagger}_{1s}a_{1s}a^{\dagger}_{1,-s}a_{1,-s}\; , 
\label{c1}
\end{eqnarray}
where $s=\pm 1/2$, and $U$ is the Coulomb repulsion energy.

Writing down the many-body wave function, $|\Psi (t)\rangle$, in 
the occupation number representation, just as in
Eq.~(\ref{wf}), and then substituting it into  
the Schr\"odinger equation $i|\dot\Psi (t)\rangle ={\cal H}|\Psi (t)\rangle$,
we find a system of coupled equations 
for the amplitudes $b(t)$ 
\begin{mathletters}
\label{c2}
\begin{eqnarray}
& &E \tilde{b}_{0}(E) - \sum_l \Omega_{l}\left [\tilde{b}_{\uparrow l}(E)
+\tilde{b}_{\downarrow l}(E)\right ]=i
\label{c2a}\\
&(&E + E_{l} - E_1) \tilde{b}_{\uparrow l}(E) - \Omega_{l}\tilde{b}_{0}(E) - 
\sum_{l'} \Omega_{l'}\tilde{b}_{\uparrow \downarrow ll'}(E)-
\sum_r \Omega_{r}\tilde{b}_{lr}(E)=0
\label{c2b}\\
&(&E+E_l-E_r)\tilde{b}_{lr}(E)- \Omega_{r}\tilde{b}_{\uparrow l}(E)-
\sum_{l'} \Omega_{l'}\left [\tilde{b}_{\uparrow l'}(E)
+\tilde{b}_{\downarrow l'}(E)\right ]=0
\label{c2c}\\ 
&(&E + E_{l} + E_{l'}- 2E_1-U) \tilde{b}_{\uparrow \downarrow ll'}(E) 
      -\Omega_{l'} \tilde{b}_{\uparrow l}(E)-\Omega_{l} 
\tilde{b}_{\downarrow l'}(E)\nonumber\\
& &~~~~~~~~~~~~~~~~~~~~~~~~~~~~~~~~~~~~~~~~~~~~~~~~~~~
-\sum_{r}\Omega_{r} \left [\tilde{b}_{\uparrow ll'r}(E)
+\tilde{b}_{\downarrow ll'r}(E)\right ]=0
\label{c2d}\\
& &\cdots\cdots\cdots\cdots\cdots
\nonumber
\end{eqnarray}
\end{mathletters}
In order to shorten notations we eliminated the index (1) of the level 
$E_1$ in the amplitudes $b$, 
so that $\tilde b_{\uparrow(\downarrow )\ldots}(t)$ 
denotes the probability amplitude 
to find one electron inside the well with spin up (down), and  
the  amplitude $\tilde b_{\uparrow\downarrow\ldots}(t)$ is the probability
amplitude to find two electrons inside the well. 

Eqs. (\ref{c2}) can be simplified 
by using the same procedure as described in the previous section. 
For instance, by substituting  $\tilde{b}_{lr}$ from Eq.~(\ref{c2c}) and 
$\tilde{b}_{\uparrow \downarrow ll'}$ from 
Eq.~(\ref{c2d}) into Eq.~(\ref{c2b}), and neglecting the ``cross terms'' 
on the grounds of the same arguments as 
in the analysis of Eq.~(\ref{exam})), we obtain 
\begin{equation} 
\left [E+E_l-E-\int_{-\infty}^{E_F^L}
\frac{\rho_L(E_{l'})\Omega^2(E_{l'})dE_{l'}}
{E + E_{l} + E_{l'}- 2E_1-U}
-\int_{E_F^R}^{\infty}\frac{\rho_R(E_{r})\Omega^2(E_{r})dE_{r}}
{E+E_l-E_r}\right ]\tilde b_{\uparrow l}(E)=0
\label{cc2} 
\end{equation} 
Since $E_l\sim E_1$, the singular parts of the integrals in (\ref{cc2}) are
respectively
$\;\,-i\Theta (E_F^L+E-E_1+U)\,\Gamma'_L/2$ and 
$\;\,-i\Theta (E+E_1-E_F^R)\,\Gamma_R/2$, where 
\begin{equation} 
\Gamma_{L(R)}=2\pi\rho_{L(R)}(E_1)|\Omega_{L(R)}(E_1)|^2, \;\;\;\;\
\Gamma_{L(R)}'=2\pi\rho_{L(R)}(E_1+U)|\Omega_{L(R)}(E_1+U)|^2.
\label{cc3} 
\end{equation}
Here $\rho_{L(R)}$ is the spin up or spin down density of states 
in the emitter (collector), 
$\rho_{L(R)}\equiv\rho_{L(R)\uparrow}=\rho_{L(R)\downarrow}$.
As in the previous section, we assume the resonance level being  
deeply inside the band, $E_F^R\ll E_1\ll E_F^L$. If,  
in addition, $E_1+U\ll E_F^L$, the theta-function in the singular 
parts of the integrals in (\ref{cc2}) can be replaced by one. 
In the opposite case, $E_1+U\gg E_F^L$, the corresponding singular part 
is zero. 

Proceeding this way with the other equations of the 
system (\ref{c2}), we finally obtain 
\begin{mathletters}
\label{c3}
\begin{eqnarray}
& &(E +i\Gamma_L)\tilde{b}_{0}(E)=i
\label{c31}\\
&(&E + E_{l} - E_1+i\Gamma_L'/2+i\Gamma_R/2) \tilde{b}_{\uparrow l}(E) - 
\Omega_{l}\tilde{b}_{0}(E) =0
\label{c32}\\
&(&E+E_l-E_r+i\Gamma_L)\tilde{b}_{lr}(E)- \Omega_{r}\tilde{b}_{\uparrow l}(E)=0
\label{c23}\\ 
&(&E + E_{l} + E_{l'}- 2E_1-U+i\Gamma_R') 
\tilde{b}_{\uparrow\downarrow ll'}(E) 
      -\Omega_{l} \tilde{b}_{\downarrow l'}(E)+
\Omega_{l'} \tilde{b}_{\uparrow l}(E)=0
\label{c34}\\
& &\cdots\cdots\cdots\cdots\cdots
\nonumber
\end{eqnarray}
\end{mathletters} 

Eqs. (\ref{c3}) can be transformed   
into equations for the density matrix of the ``device'' by using the method  
of the previous section. Since the algebra remains essentially  
the same, we give only 
the final equations for the diagonal density matrix elements
$\sigma_{aa}^{(n)}(t)$, $\sigma_{bb\uparrow}^{(n)}(t)$,  
$\sigma_{bb\downarrow}^{(n)}(t)$ and $\sigma_{cc}^{(n)}(t)$. These are  
the probabilities to find: a) no electrons inside 
the well; b) one electron with spin up (down) inside the well, and   
c) two electrons inside the well, respectively.    
The index $n$ denotes the number of 
electrons accumulated in the collector. We obtain   
\begin{mathletters}
\label{c4}
\begin{eqnarray}
\dot\sigma_{aa}^{(n)} & = &-2\Gamma_L\sigma_{aa}^{(n)}
+\Gamma_R\sigma_{bb\uparrow}^{(n-1)}+\Gamma_R\sigma_{bb\downarrow}^{(n-1)}
\label{c4a}\\
\dot\sigma_{bb\uparrow}^{(n)} & = &-(\Gamma'_L+\Gamma_R) 
\sigma_{bb\uparrow}^{(n)}+\Gamma_L\sigma_{aa}^{(n)}
+\Gamma'_R\sigma_{cc}^{(n-1)}
\label{c4b}\\
\dot\sigma_{bb\downarrow}^{(n)} & = &-(\Gamma'_L+\Gamma_R) 
\sigma_{bb\downarrow}^{(n)}+\Gamma_L\sigma_{aa}^{(n)}
+\Gamma'_R\sigma_{cc}^{(n-1)}
\label{c4c}\\
\dot\sigma_{cc}^{(n)} & = &-2\Gamma'_R\sigma_{cc}^{(n)}
+\Gamma'_L\sigma_{bb\uparrow}^{(n)}+\Gamma'_L\sigma_{bb\downarrow}^{(n)}
\label{c4d}
\end{eqnarray}
\end{mathletters} 
These rate equations look as a generalization of the rate
equations (\ref{aandb}), if one allows the well to be occupied by two 
electrons.
The Coulomb repulsion leads merely to a modification of the 
corresponding rates $\Gamma\to\Gamma'$, due to increase of the 
two electron energy. 

Summing up the partial probabilities we obtain 
for the total probabilities, 
$\sigma(t)=\sum_n\sigma^{(n)}(t)$, the following equations:    
\begin{mathletters}
\label{c5}
\begin{eqnarray}
\dot\sigma_{aa} & = &-2\Gamma_L\sigma_{aa}
+\Gamma_R\sigma_{bb\uparrow}+\Gamma_R\sigma_{bb\downarrow}
\label{c5a}\\
\dot\sigma_{bb\uparrow} & = &-(\Gamma'_L+\Gamma_R) 
\sigma_{bb\uparrow}+\Gamma_L\sigma_{aa}
+\Gamma'_R\sigma_{cc}
\label{c5b}\\
\dot\sigma_{bb\downarrow} & = &-(\Gamma'_L+\Gamma_R) 
\sigma_{bb\downarrow}+\Gamma_L\sigma_{aa}
+\Gamma'_R\sigma_{cc}
\label{c5c}\\
\dot\sigma_{cc} & = &-2\Gamma'_R\sigma_{cc}
+\Gamma'_L\sigma_{bb\uparrow}+\Gamma'_L\sigma_{bb\downarrow},
\label{c5d}
\end{eqnarray}
\end{mathletters}
and for the current  
\begin{equation}
I(t)=\sum_n n[\dot\sigma^{(n)}(t)]=
e\Gamma_R\left [\sigma_{bb\uparrow}(t)+\sigma_{bb\downarrow}(t)\right ]
+2e\Gamma'_R\sigma_{cc}(t)
\label{c6}
\end{equation}

Eqs. (\ref{c5}), (\ref{c6}) can be solved most easily for dc 
current, $I=I(t\to\infty )$. In this case $\dot\sigma =0$, and 
Eqs. (\ref{c5}) turn into the system of linear algebraic equations. 
One also finds from Eq.~(\ref{c5}) that  
$\sigma_{aa}+\sigma_{bb\uparrow}+\sigma_{bb\downarrow}+\sigma_{cc}=1$. 
The latter implies that dc current 
does not depend on the initial conditions. Finally we obtain  
\begin{equation}
I/e=\frac{2\Gamma_L\Gamma'_R(\Gamma'_L+\Gamma_R)}{\Gamma_L\Gamma'_L+
2\Gamma_L\Gamma'_R+\Gamma_R\Gamma'_R}
\label{c7}
\end{equation}

If $E_1\ll E_F^L\ll E_1+U$, one finds from Eq.~(\ref{cc2}) that 
$\Gamma'_L=0$, so that the state with two electrons inside the 
well is not available. In this case one obtains 
from Eq.~(\ref{c7}) for the dc current
\begin{equation}
I/e=\frac{2\Gamma_L\Gamma_R}{2\Gamma_L+\Gamma_R}
\label{c8}
\end{equation}
It is interesting to note that this result 
is different from Eq.~(\ref{a5}), although 
in both cases only one electron can occupy the well. 
However, if the Coulomb repulsion effect is small, i.e.  
$\Gamma'_{L,R}=\Gamma_{L,R}$, Eq.~(\ref{c7}) does produce the same result as  
Eq.~(\ref{a5}), provided the density of states is doubled 
due to the spin degrees of freedom. 

One can also consider the case when 
the Fermi level in the right reservoir $E_F^R$ lies above the resonance 
level $E_1$, but below $E_1+U$, so that $\Gamma_R=0$, Eq.~(\ref{cc2}).
Then the resonant transitions of electrons from the left to the 
right reservoirs can go only through the state with two
electrons inside the well. Using Eq.~(\ref{c7}) one finds 
for the dc current
\begin{equation}
I/e=\frac{2\Gamma'_L\Gamma'_R}{\Gamma'_L+2\Gamma'_R},
\label{c9}
\end{equation}   
which coincides with the result found by Glazmann and Matveev\cite{glaz}. 

\section{Double-well structure}
\subsection{Non-interacting electrons.}
Now we turn to the coherent case of resonant tunneling.  
Let us  consider the coupled-well structure, shown in Fig. 2. 
We assume that both levels $E_{1,2}$ are 
inside the band, i.e. $E_F^R\ll E_{1},E_{2}\ll E_F^L$.
In order to make our derivation as clear as possible, 
we begin with the case of no spin degrees of freedom and no Coulomb 
interaction.
The tunneling Hamiltonian for this system is
\begin{eqnarray}
{\cal H} &=& \sum_l E_{i}a^{\dagger}_{l}a_{l} +
              E_1 a_1^{\dagger}a_{1} + E_2 a_2^{\dagger}a_{2}+ 
              \sum_r E_{r}a^{\dagger}_{r}a_{r} 
\nonumber\\
          & & +\Omega_0(a_1^{\dagger}a_{2}+a_2^{\dagger}a_{1})
              + \sum_l \Omega_{l}(a^{\dagger}_{l}a_1 +
              a^{\dagger}_{1}a_{l})
              + \sum_r \Omega_{r}(a^{\dagger}_{r}a_2 +
              a^{\dagger}_{2}a_{r})\;.
\label{b1}
\end{eqnarray}
where $a_{1,2}^{\dagger}$, $a_{1,2}$ are creation and annihilation operators 
for an electron in the first or the second well, respectively. All the 
other notations are taken from Sect. 2.  
The many-body wave function describing this system can be written 
in the occupation number representation as 
\begin{eqnarray}
|\Psi (t)\rangle & = & \left [ b_0(t) + \sum_l b_{1l}(t)
     a_{1}^{\dagger}a_{l} + \sum_{l,r} b_{lr}(t)a_{r}^{\dagger}a_{l}\right.
      \nonumber\\
      &+& \left. \sum_l b_{2l}(t)a_{2}^{\dagger}a_{l} 
           +\sum_{ll'} b_{12ll'}(t)a_{1}^{\dagger}a_{2}^{\dagger}a_{l}a_{l'} 
           +\sum_{l<l',r} b_{1ll'r}(t)
           a_{1}^{\dagger}a_{r}^{\dagger}a_{l}a_{l'}
           + \ldots \right ] |0\rangle, 
\label{b2}
\end{eqnarray}
Substituting Eq.~(\ref{b2}) into the Shr\"odinger equation with the 
Hamiltonian (\ref{b1}) and performing the Laplace transform, 
we obtain an infinite set of the coupled equations for the 
amplitudes $\tilde b(t)$:
\begin{mathletters}
\label{b3}
\begin{eqnarray}
& &E \tilde{b}_{0}(E) - \sum_l \Omega_{l}\tilde{b}_{1l}(E)=i
\label{b3a}\\
&(&E + E_{l} - E_1) \tilde{b}_{1l}(E) - \Omega_{l}
      \tilde{b}_0(E) -\Omega_{0}\tilde{b}_{2l}(E)=0
\label{b3b}\\
&(&E + E_{l} - E_2) \tilde{b}_{2l}(E) -
      \Omega_0 \tilde{b}_{1l}(E) - 
      \sum_{l'} \Omega_{l'}\tilde{b}_{12ll'}(E)-
      \sum_r \Omega_{r}\tilde{b}_{rl}(E)=0
\label{b3c}\\
&(&E + E_{l} + E_{l'} - E_1 - E_2) \tilde{b}_{12ll'}(E)-
\Omega_{l'} \tilde{b}_{2l}(E)+
\Omega_{l} \tilde{b}_{2l'}(E)-
\sum_{r} \Omega_{r}\tilde{b}_{1ll'r}(E)=0
\label{b3d}\\
& &\cdots\cdots\cdots\cdots\cdots 
\nonumber
\end{eqnarray}
\end{mathletters}
Using exactly the same procedure as in the previous sections, Eq.~(\ref{exam}),
we transform Eqs. (\ref{b3}) into the following set of equations:
\begin{mathletters}
\label{b4}
\begin{eqnarray}
&& (E + i\Gamma_L/2) \tilde{b}_{0}(E)=i
\label{b4a}\\
&& (E + E_{l} - E_1) \tilde{b}_{1l}(E)
      - \Omega_{l} \tilde{b}_{0}(E)-\Omega_0\tilde b_{2l}(E)=0
\label{b4b}\\ 
&& (E + E_{l} - E_2+i\Gamma_L/2+ i\Gamma_R/2) \tilde{b}_{2l}(E) -
      \Omega_{0} \tilde{b}_{1l}(E)=0
\label{b4c}\\
&& (E + E_{l} + E_{l'} - E_1 - E_2 + i\Gamma_R/2) 
       \tilde{b}_{12ll'}(E) - 
       \Omega_{l'} \tilde{b}_{2l}(E) +\Omega_{l} \tilde{b}_{l'r}(E)=0
\label{b4d}\\
& &\cdots\cdots\cdots\cdots\cdots 
\nonumber
\end{eqnarray}
\end{mathletters}

The amplitudes $b(t)$ determine the density submatrix of 
the system, $\sigma_{ij}^{(n)}$, in the corresponding Fock space:
$|a\rangle$ -- the levels $E_{1,2}$ are empty,
$|b\rangle$ -- the level $E_1$ is occupied, 
$|c\rangle$ -- the level $E_2$ is occupied,
$|d\rangle$ -- the both level $E_{1,2}$ are occupied;  
the index $n$ denotes the number of electrons in 
the collector. The matrix elements of the density 
matrix of the ``device'' can be written as  
\begin{mathletters}
\label{b5}
\begin{eqnarray}
\sigma_{aa}=\sum_n\sigma_{aa}^{(n)} 
&\equiv& |b_{0}(t)|^2 + \sum_{l,r} |b_{lr}(t)|^2
            + \sum_{l<l',r<r'} |b_{ll'rr'}(t)|^2 + \ldots
\label{b5a}\\
\sigma_{bb} =\sum_n\sigma_{bb}^{(n)}&\equiv& \sum_l |b_{1l}(t)|^2 +
             \sum_{l<l',r} |b_{1ll'r}(t)|^2 
             + \sum_{l<l'<l'',r<r'} |b_{1ll'l''rr'}(t)|^2 + \ldots
\label{b5b}\\
\sigma_{cc} =\sum_n\sigma_{cc}^{(n)}&\equiv& \sum_l |b_{2l}(t)|^2 +
             \sum_{l<l',r} |b_{2ll'r}(t)|^2 
             + \sum_{l<l'<l'',r<r'} |b_{2ll'l''rr'}(t)|^2 + \ldots
\label{b5c}\\
\sigma_{dd} =\sum_n\sigma_{dd}^{(n)}&\equiv& \sum_{l<l'} |b_{12ll'}(t)|^2 +
             \sum_{l<l'<l''<l''',r<r'} |b_{12ll'l''l'''rr'}(t)|^2 
             + \ldots
\label{b5d}\\
\sigma_{bc} =\sum_n\sigma_{bc}^{(n)}&\equiv& \sum_l b_{1l}(t)b_{2l}^*(t) +
             \sum_{l<l',r} b_{1ll'r}(t) b_{2ll'r}^*(t)+ \ldots
\label{b55}  
\end{eqnarray}
\end{mathletters}

Now we transform Eqs. (\ref{b4}) into 
differential equations for $\sigma^{(n)}(t)$.
Consider for instance the term $\sigma_{bb}^{(0)}=\sum_l |b_{1l}(t)|^2$, 
Eq.~(\ref{b5b}), where the amplitudes $b_{1l}$ are determined 
by Eq.~(\ref{b4b}). Multiplying Eq.~(\ref{b4b}) by $\tilde{b}^*_{1l}(E')$ 
and subtracting the complex conjugate equation with $E\leftrightarrow E'$,
we find
\newpage 
\begin{eqnarray}
\sum_l(E'-E)\tilde b_{1l}(E)\tilde b^*_{1l}(E')-\sum_l\Omega_l
&[&\tilde b^*_{0}(E')\tilde b_{1l}(E)-\tilde b_{0}(E)\tilde b^*_{1l}(E')]
\nonumber\\
&-&\Omega_0\sum_l
[\tilde b^*_{2l}(E')\tilde b_{1l}(E)-\tilde b_{2l}(E)\tilde b^*_{1l}(E')]=0
\label{b7}
\end{eqnarray}
After applying the inverse Laplace transform, Eq.~(\ref{invlap}), 
the first term in this equation 
becomes $-i\dot\sigma_{bb}^{(0)}(t)$. 
Next, substituting
\begin{equation}
\tilde{b}_{1l}(E)=\frac{
      \Omega_{l} \tilde{b}_{0}(E)+\Omega_0\tilde b_{2l}(E)}{E + E_{l} - E_1}
\label{b6}
\end{equation}
from Eq.~(\ref{b4b}) into the second term of Eq.~(\ref{b7}),  
and replacing the sum by an integral over 
$E_l$, we reduce this term to $i\Gamma_L\tilde b_0(E)\tilde b^*_0(E')$.
After the inverse Laplace transform it becomes  
$i\Gamma_L\sigma_{aa}^{(0)}(t)$. Notice that the ``cross term'',  
$\propto \Omega_0\Omega_l\tilde b_0\tilde b_{2l}$,  
does not contribute to the integral over $E_l$, since 
the poles of the integrand in the
$E_l$-variable lie on one side of the integration contour 
(cf.~the second term of Eq.~(\ref{exam})). 
The third term of Eq.~(\ref{b7}) turns to be  
$\Omega_0[\sigma_{bc}^{(0)}(t)-\sigma_{cb}^{(0)}(t)]$,
after the inverse Laplace transform. Finally we obtain a 
differential equation for the density submatrix 
element $\sigma_{bb}^{(0)}$, 
\begin{equation}
\dot\sigma_{bb}^{(0)}(t)=\Gamma_L\sigma_{aa}^{(0)}
		+i\Omega_0(\sigma_{bc}^{(0)}-\sigma_{cb}^{(0)}). 
\label{bb7}
\end{equation}
In contrast to the rate equations of the previous sections,
the diagonal matrix element $\sigma_{bb}$ is coupled with the 
off-diagonal density matrix element $\sigma_{bc}$. 

The corresponding differential equation for $\sigma_{bc}$ can 
be easily obtained by multiplying Eq.~(\ref{b4b}) by $\tilde b_{2l}^*(E')$ 
with subsequent 
subtracting the complex conjugated Eq.~(\ref{b4c}), multiplied by 
$\tilde b_{1l}$. Afterwords by integrating over $E_l$ we obtain
\begin{equation}
\dot\sigma_{bc}^{(0)}  =  i(E_2-E_1)\sigma_{bc}^{(0)}+
i\Omega_0(\sigma_{bb}^{(0)}-\sigma_{cc}^{(0)})
-\frac{1}{2}(\Gamma_L+\Gamma_R)\sigma_{bc}^{(0)}\;.
\label{bb8}
\end{equation}

Eventually we arrive to the following set of equations for $\sigma^{(n)}$
\begin{mathletters}
\label{b8}
\begin{eqnarray}
\dot\sigma_{aa}^{(n)} & = & -\Gamma_L\sigma_{aa}^{(n)}
+\Gamma_R\sigma_{cc}^{(n-1)}\;,
\label{b8a}\\
\dot\sigma_{bb}^{(n)} & = & \Gamma_L\sigma_{aa}^{(n)}
+\Gamma_R\sigma_{dd}^{(n-1)}+i\Omega_0(\sigma_{bc}^{(n)}-\sigma_{cb}^{(n)})\;,
\label{b8b}\\
\dot\sigma_{cc}^{(n)} & = & -\Gamma_R\sigma_{cc}^{(n)}
-\Gamma_L\sigma_{cc}^{(n)}-i\Omega_0(\sigma_{bc}^{(n)}-\sigma_{cb}^{(n)})\;,
\label{b8c}\\
\dot\sigma_{dd}^{(n)} & = & -\Gamma_R\sigma_{dd}^{(n)}
+\Gamma_L\sigma_{cc}^{(n)}\;,
\label{b8d}\\
\dot\sigma_{bc}^{(n)} & = & i(E_2-E_1)\sigma_{bc}^{(n)}+
i\Omega_0(\sigma_{bb}^{(n)}-\sigma_{cc}^{(n)})
-\frac{1}{2}(\Gamma_L+\Gamma_R)\sigma_{bc}^{(n)}\;.
\label{b8e}
\end{eqnarray}
\end{mathletters}
Using Eqs. (\ref{b8}) we can find the charge accumulated in the collector,
$N_R(t)$, and subsequently, the total current, $e\dot N(t)$, as given by  
\begin{equation}
I(t)/e = \dot N(t)=\sum_n n\left [\dot\sigma_{aa}^{(n)}(t)
+\dot\sigma_{bb}^{(n)}(t)
+\dot\sigma_{cc}^{(n)}(t)+\dot\sigma_{dd}^{(n)}(t)\right ]
=\Gamma_R \left [\sigma_{cc}(t) + \sigma_{dd}(t)\right ] 
\label{b9}
\end{equation}
As in the previous examples, the current is    
proportional to the total probability 
of finding an electron in the well adjacent to the 
right reservoir. The off-diagonal elements of the density matrix do
not appear in Eq.~(\ref{b9}). 

Summing up over $n$ in Eqs. (\ref{b8}), we obtain the system of differential 
equations for the density matrix elements of the device
\begin{mathletters}
\label{b10}
\begin{eqnarray}
\dot\sigma_{aa} & = & -\Gamma_L\sigma_{aa}
+\Gamma_R\sigma_{cc}\;,
\label{b10a}\\
\dot\sigma_{bb} & = & \Gamma_L\sigma_{aa}
+\Gamma_R\sigma_{dd}+i\Omega_0(\sigma_{bc}-\sigma_{cb})\;,
\label{b10b}\\
\dot\sigma_{cc} & = & -\Gamma_R\sigma_{cc}
-\Gamma_L\sigma_{cc}-i\Omega_0(\sigma_{bc}-\sigma_{cb})\;,
\label{b10c}\\
\dot\sigma_{dd} & = & -\Gamma_R\sigma_{dd}
+\Gamma_L\sigma_{cc}\;,
\label{b10d}\\
\dot\sigma_{bc} & = & i(E_2-E_1)\sigma_{bc}+
i\Omega_0(\sigma_{bb}-\sigma_{cc})
-\frac{1}{2}(\Gamma_L+\Gamma_R)\sigma_{bc}\;,
\label{b10e}
\end{eqnarray}
\end{mathletters}
Eqs.~(\ref{b10})) resemble the optical Bloch equations\cite{bloch}. 
Note that the coupling with the reservoirs produces purely 
negative contribution into the {\em non-diagonal} 
matrix element's dynamic equation,  
Eq.~(\ref{b10e}), thus causing damping of this matrix element.

Eqs.~(\ref{b10}) are solved most easily for the 
stationary current, $I=I(t\to\infty )$. Using
$\sigma_{aa}+\sigma_{bb}+\sigma_{cc}+\sigma_{dd}=1$, we obtain    
\begin{equation}
I/e=\left (\frac{\Gamma_L\Gamma_R}{\Gamma_L+\Gamma_R}\right )
\frac{\Omega_0^2}{\Omega_0^2+\Gamma_L\Gamma_R/4+\epsilon^2\Gamma_L\Gamma_R/
(\Gamma_L+\Gamma_R)^2}\;,
\label{b11}
\end{equation}
where $\epsilon =E_2-E_1$. This result coincides with the one found 
in the framework of one-electron approach \cite{g,sok}. 

\subsection{Coulomb blockade.}
The extension of the rate equations (\ref{b10}) for the case of spin and 
Coulomb interaction is done exactly in the same way as in Sect 3.
Here also the rate equations for the device density matrix are
obtained only for $E_{1,2}+U$ being inside or outside the band, but not
close to the band edges ($E_F^R\ll E_{1,2}+U\ll E_F^L$ or
$E_{1,2}+U\gg E_F^L$). 
Eventually we arrive to the rate equations of type Eqs.~(\ref{b10}), 
but with the number of the available states of the device changed 
due to additional (spin) degrees of freedom and Coulomb blockade
restrictions. The Coulomb repulsion  
manifests itself also in a modification of the
transition amplitude $\Omega$ and the rates $\Gamma$'s, Eq.~(\ref{cc3}). 

In the case of large Coulomb repulsion, some of electron states 
of the device are outside the band (the Coulomb blockade). As a result,  
the number of the equations is reduced.   
Consider, for instance, the situation where the Coulomb interaction $U$ of two 
electrons in the same well so large that $E_{1,2}+U\gg E_F^L$, but
the Coulomb repulsion of two electrons in different wells, $\bar U$, is 
much smaller, so that $E_{1,2}+\bar U\ll E_F^L$. 
Then the state of two electrons in the same well is not available, but
two electrons can occupy different wells. In this case the rate 
equations for the corresponding density matrix elements of the device are    
\begin{mathletters}
\label{b12}
\begin{eqnarray}
& &\dot\sigma_{aa} = -2\Gamma_L\sigma_{aa}
+\Gamma_R(\sigma_{cc\uparrow}+\sigma_{cc\downarrow})\;,
\label{b12a}\\
& &\dot\sigma_{bb\uparrow} = \Gamma_L\sigma_{aa}
-\Gamma'_R(\sigma_{dd\uparrow\uparrow}+\sigma_{dd\uparrow\downarrow})
+i\Omega_0(\sigma_{bc\uparrow}-\sigma_{cb\uparrow})\;,
\label{b12b}\\
& &\dot\sigma_{cc\uparrow} = -\Gamma_R\sigma_{cc\uparrow}
-2\Gamma'_L\sigma_{cc\uparrow}
-i\Omega_0(\sigma_{bc\uparrow}-\sigma_{cb\uparrow})\;,
\label{b12c}\\
& &\dot\sigma_{dd\uparrow\uparrow} = -\Gamma'_R\sigma_{dd\uparrow\uparrow}
+\Gamma'_L\sigma_{cc\uparrow}\;,
\label{b12d}\\
& &\dot\sigma_{bc\uparrow} = i(E_2-E_1)\sigma_{bc\uparrow}+
i\Omega_0(\sigma_{bb\uparrow}-\sigma_{cc\uparrow})
-\frac{1}{2}(2\Gamma'_L+\Gamma_R)\sigma_{bc\uparrow}\;,
\label{b12e}
\end{eqnarray}
\end{mathletters}  
where $\Gamma'_{L(R)}=2\pi\rho_{L(R)}(E_1+\bar U)
|\Omega_{L(R)}(E_1+\bar U)|^2$. Here for the shortness 
we wrote only the equations for the ``spin up'' component of 
the density matrix. The same equations are obtained for the 
``spin down'' components of the density matrix. The total current is  
\begin{equation}
I/e=\Gamma_R(\sigma_{cc\uparrow}+\sigma_{cc\downarrow})+
\Gamma'_R(\sigma_{dd\uparrow\uparrow}+\sigma_{dd\uparrow\downarrow}+
\sigma_{dd\downarrow\uparrow}+\sigma_{dd\downarrow\downarrow}).
\label{b13}
\end{equation}

It is quite clear that the ``spin up'' and ``spin down'' 
components of the density matrix are equal, i.e.  
$\sigma_{bb\uparrow} = \sigma_{bb\downarrow}
= \sigma_{bb}$, the same holding for $\sigma_{cc}$, $\sigma_{dd}$ 
components. Therefore 
Eqs. (\ref{b12}), (\ref{b13}) can be rewritten as 
\begin{mathletters}
\label{b14}
\begin{eqnarray}
\dot\sigma_{aa} & = & -2\Gamma_L\sigma_{aa}
+2\Gamma_R\sigma_{cc}\;,
\label{b14a}\\
\dot\sigma_{bb} & = & \Gamma_L\sigma_{aa}
+2\Gamma'_R\sigma_{dd}+i\Omega_0(\sigma_{bc}-\sigma_{cb})\;,
\label{b14b}\\
\dot\sigma_{cc} & = & -\Gamma_R\sigma_{cc}
-2\Gamma'_L\sigma_{cc}-i\Omega_0(\sigma_{bc}-\sigma_{cb})\;,
\label{b14c}\\
\dot\sigma_{dd} & = & -\Gamma'_R\sigma_{dd}
+\Gamma'_L\sigma_{cc}\;,
\label{b14d}\\
\dot\sigma_{bc} & = & i(E_2-E_1)\sigma_{bc}+
i\Omega_0(\sigma_{bb}-\sigma_{cc})
-\frac{1}{2}(2\Gamma'_L+\Gamma_R)\sigma_{bc}\;,
\label{b14e}
\end{eqnarray}
\end{mathletters}
and 
\begin{equation}
I/e=2\Gamma_R\sigma_{cc}+4\Gamma'_R\sigma_{dd}
\label{b15}
\end{equation}

Using $\sigma_{aa}+2\sigma_{bb}+2\sigma_{cc}+4\sigma_{dd}=1$ we obtain for 
the dc current 
\begin{equation}  
I/e=\left (\frac{2\Gamma_L\Gamma'_R}{2\Gamma'_L+\Gamma_R}\right )
\frac{\Omega_0^2}{\displaystyle 4\Omega_0^2\frac{\Gamma_L\Gamma'_L+
\Gamma_L\Gamma'_R+\Gamma_R\Gamma'_R/4}
{(2\Gamma'_L+\Gamma_R)^2}
+\frac{\Gamma_L\Gamma'_R}{2}+\epsilon^2\frac{2\Gamma_L\Gamma'_R}
{(2\Gamma'_L+\Gamma_R)^2}}\;,
\label{b16}
\end{equation}
where $\epsilon =E_2-E_1$. Notice that the  
current (\ref{b16}) differs from that given by Eq.~(\ref{b11}) 
even for $\Gamma'_L=\Gamma_L$ and  $\Gamma'_R=\Gamma_R$,
despite the fact that in the both cases only one electron can occupy 
each of the wells. 

It is interesting to compare our result with that of Stoof and 
Nazarov\cite{naz1} for the case of strong Coulomb repulsion 
between two electrons in different wells ($E_{1,2}+\bar U\gg E_F^L$), 
where only one  
electron can be found inside the system. It corresponds 
to $\Gamma'_L=0$. In this case the dc current given by  Eq.~(\ref{b16}) is
\begin{equation}  
I/e=\frac{\Gamma_R\Omega_0^2}{\displaystyle\Omega_0^2(2+\Gamma_R/2\Gamma_L)+
\Gamma_R^2/4+\epsilon^2}. 
\label{b17}
\end{equation} 
This result is slightly different from that obtained by Stoof and 
Nazarov (by the factor two in front of $\Gamma_L$). The difference stems from  
the account of spin components in the rate equations, which has 
not been done in\cite{naz1}. 

\section{Inelastic processes}
As an example of a system with coherent tunneling  
accompanied by inelastic scattering, let us consider
the coupled-dot structure shown in Fig.~3.  
In this system a resonant current flows
due to inelastic transition from the upper to the lower level
in the left well. For simplicity, we restrict ourselves to non-interacting
spin-less electrons. The Coulomb interaction and the spin effects
can be accounted for precisely in the same way as we did in the previous
sections, namely by allowing for states with doubly occupied levels 
(excluding states violating Coulomb restrictions) and modifying transition
amplitudes and inelastic rates. 

The tunneling Hamiltonian of the system has the following structure
\begin{eqnarray}
{\cal H} & = & \sum_l E_{l}a^{\dagger}_{l}a_{l} +
              E_1 a_1^{\dagger}a_{1} + E_2 a_2^{\dagger}a_{2}+
              E_3 a_3^{\dagger}a_{3}+
              \sum_{\alpha} E_{\alpha}^{ph}c^{\dagger}_{\alpha}c_{\alpha}+  
              \sum_r E_{r}a^{\dagger}_{r}a_{r}+
              \Omega_0(a_2^{\dagger}a_{3}+a_3^{\dagger}a_{2}) 
\nonumber\\
           &+& \sum_l \Omega_{l}(a^{\dagger}_{l}a_1 +
              a^{\dagger}_{1}a_{l})+
   \sum_\alpha \Omega_{\alpha}^{ph}(a^{\dagger}_{2}a_{1}c_{\alpha}^{\dagger} +
              a^{\dagger}_{1}a_{2}c_{\alpha})+
              \sum_r \Omega_{r}(a^{\dagger}_{r}a_3+ 
              a^{\dagger}_{3}a_{r}).
\label{e1}
\end{eqnarray} 
Here the subscript $\alpha$ enumerates the states in the phonon bath
and $\Omega_{\alpha}^{ph}$ is the corresponding coupling. The 
many particle time-dependent wave function of the system is
\begin{eqnarray}
|\Psi (t)\rangle & = & \left [ b_0(t) + \sum_l b_{1l}(t)
     a_{1}^{\dagger}a_{l} + \sum_{l,\alpha} b_{2l\alpha}(t)a_{2}^{\dagger}
a_{l}c_{\alpha}^{\dagger}+\sum_{l,\alpha} b_{3l\alpha}(t)a_{3}^{\dagger}
a_{l}c_{\alpha}^{\dagger}+\right.
      \nonumber\\
      &+& \left. \sum_{l<l',\alpha} b_{12ll'\alpha}(t)
           a_{1}^{\dagger}a_{2}^{\dagger}a_{l}a_{l'}c_{\alpha}^{\dagger}
          +\sum_{l<l',\alpha} b_{13ll'\alpha}(t)
           a_{1}^{\dagger}a_{3}^{\dagger}a_{l}a_{l'}c_{\alpha}^{\dagger}
           + \ldots \right ] |0\rangle. 
\label{e2}
\end{eqnarray} 
Repeating the procedure of the previous sections we find the following set 
of equations for the Laplace transformed amplitudes, $\tilde b(E)$:
\begin{mathletters}
\label{e3}
\begin{eqnarray}
&& (E + i\Gamma_L/2) \tilde{b}_{0}=i
\label{e3a}\\
&& (E + E_{l} - E_1+i\Gamma_{in}/2) \tilde{b}_{1l}-\Omega_{l} \tilde b_0=0
\label{e3b}\\ 
&& (E + E_{l} - E_{\alpha}-E_2+i\Gamma_L/2) \tilde{b}_{2l\alpha} -
      \Omega_{\alpha}^{ph}\tilde b_{1l}-\Omega_{0} \tilde{b}_{3l\alpha}=0
\label{e3c}\\
&& (E + E_{l} -E_{\alpha}-E_3+i\Gamma_L/2+i\Gamma_R/2) \tilde{b}_{3l\alpha} -
      \Omega_{0} \tilde{b}_{2l\alpha}=0
\label{e3d}\\
&& (E + E_{l} + E_{l'} - E_1 - E_2 -E_{\alpha})\tilde{b}_{12ll'\alpha} - 
       \Omega_{l'} \tilde{b}_{2l\alpha}+\Omega_{l} \tilde{b}_{2l'\alpha} 
       -\Omega_{0}\tilde{b}_{13ll'\alpha}=0
\label{e3e}\\
&& (E + E_{l} + E_{l'} - E_1 - E_3 -E_{\alpha}+ i\Gamma_{in}/2
 + i\Gamma_R/2)\tilde{b}_{13ll'\alpha} \nonumber\\ 
&&~~~~~~~~~~~~~~~~~~~~~~~~~~~~~~~~~~~~~~~~~~~~~~~~~~~~~
-\Omega_{0}\tilde{b}_{12ll'\alpha}-\Omega_{l'}\tilde{b}_{3l\alpha}
+\Omega_{l}\tilde{b}_{3l'\alpha}=0
\label{e3f}\\
& &\cdots\cdots\cdots\cdots\cdots\nonumber 
\end{eqnarray}
\end{mathletters}
where $\Gamma_{in}=2\pi\rho_{ph}|\Omega^{ph}|^2$ is the partial width 
of the level $E_1$ due to phonon emission and $\rho_{ph}$ is the density of 
phonon states. 

The density matrix elements of the device is $\sigma_{ij}(t)=
\sum_n\sigma_{ij}^{(n)}(t)$, where $\sigma_{ij}^{(n)}(t)$, are 
related to the amplitudes $\tilde b(E)$ via Eq.~(\ref{invlap}).
All possible states electron states of the device are shown 
in Fig.~4. Using the previous section procedure for {\em diagonal} 
matrix elements we obtain master equations analogous to Eq.~(\ref{b10}), 
in which transitions between isolated levels $E_2$ and $E_3$ take place 
through the coupling with non-diagonal matrix elements. These equations have   
the appearance of the optical Bloch equation\cite{bloch}. However, the master 
equation for the {\em non-diagonal} matrix element, $\sigma_{ef}$,  
contains an additional term. Therefore, we present the derivation of
the master equations for ``coherences'' $\sigma_{ef}$ and 
$\sigma_{cd}$ in some detail. Consider for example  
the non-diagonal density submatrix elements
$\sigma_{cd}^{(0)}=\sum_{l,\alpha}b_{2l\alpha}(t)b^*_{3l\alpha}(t)$ and 
$\sigma_{ef}^{(0)}=\sum_{l<l',\alpha}b_{12ll'\alpha}(t)b^*_{13ll'\alpha}(t)$.
The differential equation for $\sigma_{cd}^{(0)}(t)$ 
can be obtained by multiplying Eq.~(\ref{e3c}) by $\tilde b^*_{3l\alpha}(E')$
with subsequent subtraction of the complex conjugated Eq.~(\ref{e3d}), 
multiplied by $\tilde b_{2l\alpha}(E)$. Then using Eq.~(\ref{invlap}), 
we obtain 
\begin{equation}  
\dot\sigma_{cd}^{(0)}  =  i(E_3-E_2)\sigma_{cd}^{(0)}+
i\Omega_0(\sigma_{cc}^{(0)}-\sigma_{dd}^{(0)})
-\frac{1}{2}(2\Gamma_L+\Gamma_R)\sigma_{cd}^{(0)}\;.
\label{e4}
\end{equation} 
Similarly, multiplying Eq.~(\ref{e3e}) by 
$\tilde b^*_{13ll'\alpha}(E')$ and  Eq.~(\ref{e3f}) 
by $\tilde b_{12ll'\alpha}(E)$, we find 
the differential equation for $\sigma_{ef}^{(0)}(t)$
\begin{equation}  
\dot\sigma_{ef}^{(0)}  =  i(E_3-E_2)\sigma_{ef}^{(0)}+
i\Omega_0(\sigma_{ee}^{(0)}-\sigma_{ff}^{(0)})
-\frac{1}{2}(\Gamma_{in}+\Gamma_R)\sigma_{ef}^{(0)}-i\Delta.
\label{e5}
\end{equation}    
where 
\begin{eqnarray}  
\Delta=\sum_{l<l',\alpha}\int\frac{dEdE'}{4\pi^2}\left [
\tilde{b}^*_{13ll'\alpha}(E')\Omega_{l'}\tilde b_{2l\alpha}(E)-
\tilde{b}^*_{13ll'\alpha}(E')\Omega_{l}\tilde b_{2l'\alpha}(E)\right.
\nonumber\\
\left.-\tilde{b}_{12ll'\alpha}(E)\Omega_{l'}\tilde b^*_{3l\alpha}(E')+
\tilde{b}_{12ll'\alpha}(E)\Omega_{l}\tilde b^*_{3l'\alpha}(E')\right ]
e^{i(E'-E)t}
\label{e6}
\end{eqnarray}
Substituting the amplitudes $\tilde{b}_{12ll'\alpha}$ 
from Eq.~(\ref{e3e}) and 
$\tilde{b}^*_{13ll'\alpha}$ from Eq.~(\ref{e3f}) into Eq.~(\ref{e6}),  
and replacing the sum over $l(l')$ by the corresponding integral, 
we find $-i\Delta=\Gamma_L\sigma_{cd}^{(0)}$. It implies that 
the non-diagonal density matrix $\sigma_{ef}$ given by Eq.~(\ref{e5}), 
is coupled with $\sigma_{cd}$ via a single electron transition 
from the emitter to the left well. Such a term does not appear in the Bloch 
equations, which deal with two-level systems.    

Summing up over $n$ in the rate equations for the density submatrix 
$\sigma^{(n)}_{ij}(t)$ we obtain the set of rate equations for  
the density matrix of the device 
\begin{mathletters}
\label{e8}
\begin{eqnarray}
\dot\sigma_{aa} & = & -\Gamma_L\sigma_{aa}+\Gamma_{R}\sigma_{dd}\;,
\label{e8a}\\
\dot\sigma_{bb} & = & \Gamma_{L}\sigma_{aa}-\Gamma_{in}
\sigma_{bb}+\Gamma_{R}\sigma_{ff}\;,
\label{e8b}\\
\dot\sigma_{cc} & = & \Gamma_{in}\sigma_{bb}+i\Omega (\sigma_{cd}-\sigma_{dc})
+\Gamma_R\sigma_{gg}-\Gamma_{L}\sigma_{cc}\;,
\label{e8c}\\
\dot\sigma_{dd} & = & -\Gamma_R\sigma_{dd}+i\Omega (\sigma_{dc}-\sigma_{cd})
-\Gamma_{L}\sigma_{dd}\;,
\label{e8d}\\
\dot\sigma_{ee} & = & \Gamma_L\sigma_{cc}+i\Omega (\sigma_{ef}-\sigma_{fe})
+\Gamma_{R}\sigma_{hh}\;,
\label{e8e}\\
\dot\sigma_{ff} & = & \Gamma_L\sigma_{dd}-\Gamma_{R}\sigma_{ff}
+i\Omega (\sigma_{fe}-\sigma_{ef})-\Gamma_{in}\sigma_{ff}\;,
\label{e8f}\\
\dot\sigma_{gg} & = & \Gamma_{in}\sigma_{ff}-\Gamma_{R}\sigma_{gg}
-\Gamma_L\sigma_{gg}\;,
\label{e8g}\\
\dot\sigma_{hh} & = & \Gamma_L\sigma_{gg}-\Gamma_{R}\sigma_{hh}\;,
\label{e8h}\\
\dot\sigma_{cd} & = & i(E_3-E_2)\sigma_{cd}+
i\Omega (\sigma_{cc}-\sigma_{dd})-1/2(2\Gamma_L+\Gamma_R)
\sigma_{cd}\;,
\label{e8i}\\
\dot\sigma_{ef} & = & i(E_3-E_2)\sigma_{ef}+
i\Omega (\sigma_{ee}-\sigma_{ff})-1/2(\Gamma_{in}+
\Gamma_R)\sigma_{ef}+\Gamma_L\sigma_{cd}\;,
\label{e8j}
\end{eqnarray}
\end{mathletters}
and the resonant current flowing through this system is 
$I/e=\Gamma_R[\sigma_{dd}+\sigma_{ff}+\sigma_{gg}+\sigma_{hh}]$.

\section{General case}
Now utilizing the results obtained in
the previous sections we can write the
rate equations for the general case.
These equations describing the time evolution of the density 
matrix $\sigma_{ab}(t)$ of the device are as follows:
\begin{mathletters}
\label{d1}
\begin{eqnarray}
\dot\sigma_{aa} & = &
 i\sum_{b(\neq a)}\Omega_{ab}(\sigma_{ab}-\sigma_{ba})
-\sigma_{aa} \sum_{d(\neq a)}\Gamma_{a \rightarrow d} +
 \sum_{c(\neq a)} \sigma_{cc}\Gamma_{c \rightarrow a}\;,
 \label{d1a}\\
\dot\sigma_{ab} & = & i(E_b - E_a) \sigma_{ab} +
i\left (\sum_{b'(\neq b)}\sigma_{ab'}\Omega_{b'b}
-\sum_{a'(\neq a)}\Omega_{aa'}\sigma_{a'b}\right )
\nonumber\\
    &  & -\frac{1}{2}\sigma_{ab}
                 \left ( \sum_{d (\neq a)}\Gamma_{a \rightarrow d}
                  +\sum_{d (\neq b)}\Gamma_{b \rightarrow d}\right )
                  +\frac{1}{2}\sum_{a'b'\neq ab}\sigma_{a'b'}
        \left (\Gamma_{a'\rightarrow a}+\Gamma_{b'\rightarrow b}\right ),
\label{d1b}
\end{eqnarray}
\end{mathletters}
where $\Omega_{ab}$ denote the couplings between non-orthogonal 
isolated states, as for instance between the levels
in adjacent wells, and $\sigma_{ba}=\sigma^*_{ab}$. The width $\Gamma_{a\to b}$ 
is the probability per unit time for the system to make a transition from
the state $|a\rangle$ to
the state $|b\rangle$ of the device due to the tunneling to (or from) the
reservoirs, or due to interaction with the phonon bath, or any other 
interaction, generated by a continuum state medium.  
Notice that Eq.~(\ref{d1a}) for diagonal elements has a classical 
rate equation form, 
except for the first term. This term describes transitions between 
isolated states through the coupling with non-diagonal terms. Therefore  
it is responsible for coherent (quantum) effects in the transport.

The non-diagonal matrix elements are described by Eq.~(\ref{d1b}), which 
resembles the corresponding Bloch equation, supplemented with an additional 
term. The latter appears whenever a {\em one-electron} transition converts 
the state $|a'\rangle$ into $|a\rangle$ {\em and} the state 
$|b'\rangle$ into $|b\rangle$. 
The positive sign of the additional term calls forth a suspicion
that Eq.~(\ref{d1b}) might have unbounded solutions. This is not the
case, however, since any positive contribution from the additional term
in the equation for $\dot\sigma_{ab}$ has its negative counterpart
in the equation for $\dot\sigma_{a'b'}$, which corresponds to the 
conversion $(a'b')\to (ab)$ and originates from the third
term in the rhs of Eq.~(\ref{d1b}). Moreover, the
negative contributions, corresponding to the conversions $(ab)\to (a'b)$
and $(ab)\to (ab')$ have no positive counterparts.
Therefore the coupling with continuum leads to negative total balance, and
hence, to the damping of non-diagonal matrix elements.  

The current through the mesoscopic device is the time derivative of 
the total charge accumulated in the collector. We find that 
the current is totally determined through  
the {\em diagonal} elements of the density matrix of the device by 
the following relation
\begin{equation}  
I(t)=e\sum_c\sigma_{cc}(t)\Gamma^{(c)}_R, 
\label{d2}
\end{equation} 
where $|c\rangle$ are the occupied states in the well 
adjacent to the collector, and 
$\Gamma_R^{(c)}$ is the partial width of the state $|c\rangle$
due to tunneling to the collector. 

Although the non-diagonal density-matrix elements do not enter
explicitly in Eq.~(\ref{d2}), they are coupled with 
diagonal matrix elements in the rate equations
(the first term in (\ref{d1a})), and therefore influence the resonant current.
The coupling with non-diagonal elements always appears
in the rate equation, whenever a carrier 
jumps from one to another {\em isolated} states inside the device. 
In the absence of such transition as, for
instance, in resonant tunneling through a single well, 
the diagonal and non-diagonal matrix elements 
are decoupled and the evolution of diagonal density-matrix elements is 
described by the {\em classical} rate equation.

Hence, the distinction between isolated and continuum states
becomes very essential in the description of quantum transport.
At first sight, it may seem that in a real situation such a distinction
can hardly be carried out, since
there are no pure isolated states. For instance,
a single electron state inside the device is always coupled with the
continuum states of phonons. However, the corresponding density
of states would display peaks in energy dependence, and they
can be considered as isolated states. Indeed, if we have written
equations like Eqs.~(\ref{exam}), (\ref{cc2}), etc. for such a system,
the contribution from these peaks in the integrals over continuum states
would generate a coupling with non-diagonal density matrix
elements in the rate equations, just as in a transition
between two isolated states, Eqs.~(\ref{d1}).

As an example of application of Eq.~(\ref{d1}) in the case of strong
Coulomb blockade, we consider
the system shown in Fig. 5. The wells may represent two coupled dots.
An electron tunnels from the emitter to the first well, and then
to the second well into the upper level $E_2$. After that it can either relax
inelastically into the lower level $E_3$ due to interaction with
the phonon bath, and then tunnel into the collector, or tunnel out
directly into the collector. Here $\Gamma_{in}$ and $\Gamma'_R$ are
the partial widths of the upper level, $E_2$, due to coupling to phonon
reservoir and the collector, respectively, and $\Gamma_R$ is the
width of the level $E_3$ due to coupling to the collector.
Let's assume that the Coulomb blockade prevents the system 
from being occupied by two electrons, even in different wells.   
Then there are four possible states of the device 
$|a\rangle$ -- all the levels $E_{1,2,3}$ are empty; $|b\rangle$ -- the 
level $E_1$ is occupied; $|c\rangle$ -- the level $E_2$ is occupied; 
$|d\rangle$ -- the level $E_3$ is occupied. It is clear that the
density matrix elements for an electron with spin up and spin down 
inside the system are equal, $\sigma_{bb\uparrow}= 
\sigma_{bb\downarrow}= \sigma_{bb}$, and the same holds for 
$\sigma_{cc}$ and $\sigma_{dd}$. Hence, Eqs. (\ref{d1}) can be 
written in this case as   
\begin{mathletters}
\label{d3}
\begin{eqnarray}
\dot\sigma_{aa} & = & -2\Gamma_L\sigma_{aa}
+2\Gamma'_R\sigma_{cc}+2\Gamma_R\sigma_{dd}\;,
\label{d3a}\\
\dot\sigma_{bb} & = & i\Omega_0(\sigma_{bc}-\sigma_{cb})+\Gamma_L\sigma_{aa}\;,
\label{d3b}\\
\dot\sigma_{cc} & = & -i\Omega_0(\sigma_{bc}-\sigma_{cb})
-(\Gamma'_R+\Gamma_{in})\sigma_{cc}\;,
\label{d3c}\\
\dot\sigma_{dd} & = & -\Gamma_R\sigma_{dd}+\Gamma_{in}\sigma_{cc}\;,
\label{d3d}\\
\dot\sigma_{bc} & = & i(E_2-E_1)\sigma_{bc}+
i\Omega_0(\sigma_{bb}-\sigma_{cc})
-\frac{1}{2}(\Gamma'_R+\Gamma_{in})\sigma_{bc}\;,
\label{d3e}
\end{eqnarray}
\end{mathletters}
and the dc current $I$, Eq.~(\ref{d2}), is given by
\begin{equation}  
I/e=2\sigma_{cc}\Gamma'_R+2\sigma_{dd}\Gamma_R.  
\label{d4}
\end{equation}
Using $\sigma_{aa}+2\sigma_{bb}+2\sigma_{cc}+2\sigma_{dd}=1$,
Eqs. (\ref{d3}) can be easily solved for $t\to\infty$, yielding
for the dc current  
\begin{equation}  
I/e=\left (\frac{2\Gamma_L\Gamma_R}{\Gamma_{in}+\Gamma'_R}\right )
\frac{\Omega_0^2}{\displaystyle \Omega_0^2\frac{2\Gamma_{in}\Gamma_L+
\Gamma_{in}\Gamma_R+4\Gamma_L\Gamma_R+\Gamma_R\Gamma'_R}
{(\Gamma_{in}+\Gamma'_R)^2}
+\frac{\Gamma_L\Gamma_R}{2}+\epsilon^2\frac{2\Gamma_L\Gamma_R}
{(\Gamma_{in}+\Gamma'_R)^2}}\;,
\label{d5}
\end{equation}
This result shows very peculiar dependence of the dc current of the
inelastic width $\Gamma_{in}$. One could expect, at least for
$\Gamma'_R\sim \Gamma_R$, that the current should increase when
$\Gamma_{in}$ grows. However, as follows from Eq.~(\ref{d3}), the
current $I\to 0$ for $\Gamma_{in}\to\infty$ (cf. with another
example in\cite{glp1}). In fact, such an unexpected behavior of the dc current
would always take place in the presence of coherent transitions between 
isolated states in carrier transport. For instance, it
can be traced even in a more simple case of the resonant tunneling
through a double well structure, Eq.~(\ref{b17}). One finds that
$I\to\ 0$ when $\Gamma_R\to\infty$.
This phenomenon can be understood by analyzing Eq.~(\ref{d1b}) for 
non-diagonal density matrix elements. In contrast with the rate equation 
for diagonal matrix elements, Eq.~(\ref{d1a}), the coupling with 
continuum states always leads to damping of non-diagonal matrix elements.   
Since the transport through isolated states
goes only via non-diagonal density-matrix elements, Eq.~(\ref{d1a}), 
the total current would always decrease with the growth of
the corresponding partial widths.

\section{Summary}
In this paper we have studied quantum transport in mesoscopic
systems (quantum dots) containing finite number of isolated quantum states.
Starting with the many-particle wave function in the
occupation number representation, and integrating out the continuum states,
we have found the equations of motion for the density
submatrix of the system. These equations have a form of the master (rate)
equations for diagonal density matrix element. But in addition,
non-diagonal density matrix elements, responsible for transitions between 
isolated quantum states, appear in these equations.
If, however, these transitions are generated
by a continuum states medium, the diagonal and
non-diagonal density matrix elements become decoupled, and the
quantum transport is described by classical rate equations.

It follows from our derivation that the reduction of 
many-body Schr\"odinger equation to the modified rate equations for 
density submatrix of the device can be performed only if two conditions
are met: first, the energy states of the system which carry 
the resonant transport must be inside the bias, $E_F^L-E_F^R$; second, 
the width of these states is much 
smaller than the bias. If the second condition is satisfied, but 
the resonant levels of the device are close to band edges, 
our rate equations cannot be derived. Yet, the method 
still can be used for dc current. However, when the bias is less than 
the level width, the continuum states of the reservoir cannot be integrated
out in the manner of Sect. 2, and our method cannot be applied. 
 
We have compared some of our results with ones obtained
earlier in the literature. For example, for the resonant tunneling
through a single dot we obtained the same result as
Glazman and Matveev\cite{glaz}.
In the case of resonant tunneling through double-dot structure
we found simple analytical expression for dc current under condition of
strong Coulomb repulsion inside the dots, when no more than 
one electron can occupy the dots. The obtained expression is very close to
that found in\cite{naz,naz1}. 

As an application of our equations we considered a more complicated case of
the resonant tunneling in a coupled dot system, where the inelastic
process takes place in the course of transport. It was found that
the resonant current decreases with the growth of the inelastic width. 
We found that this anomalous behavior always emerges whenever 
coherent transitions are accompanied by inelastic processes.

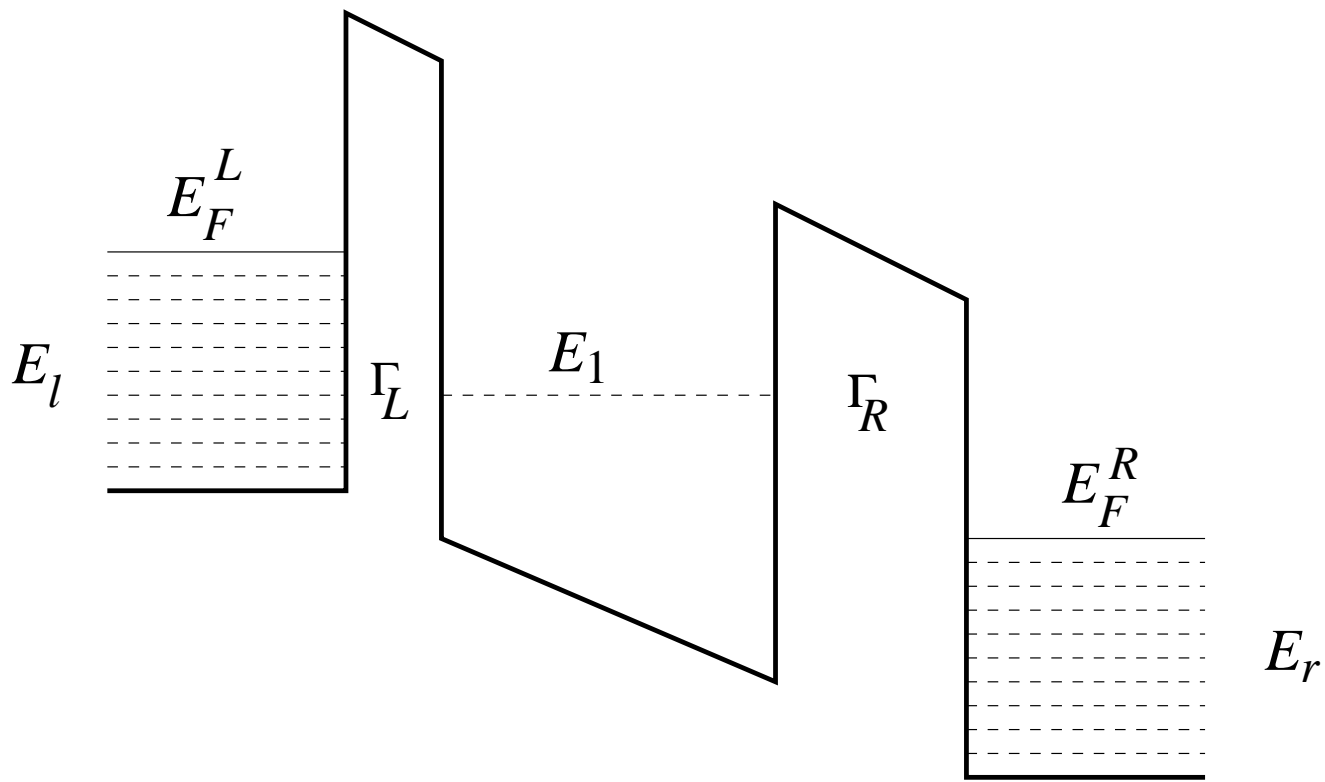
\begin{figure}[h]
\caption{Resonant transport 
through a single quantum well structure.}
\end{figure}
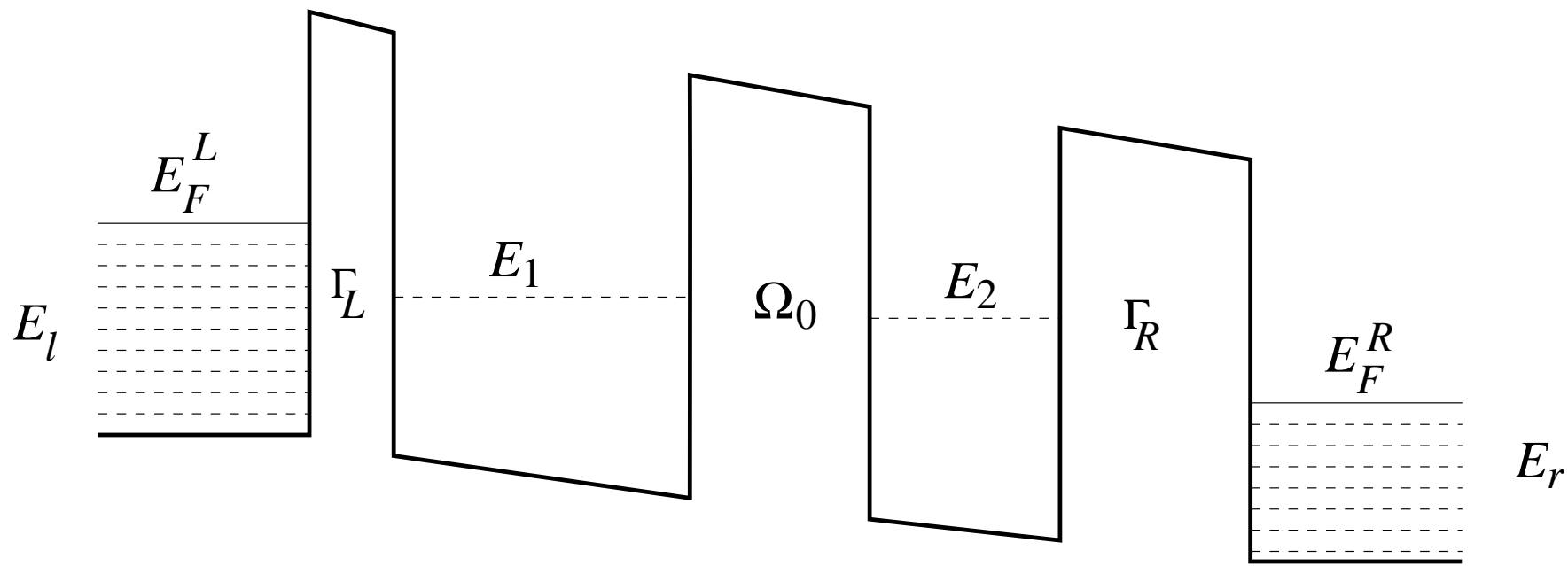
\begin{figure}[h]
\caption{Resonant transport through a double-well structure.}
\end{figure}
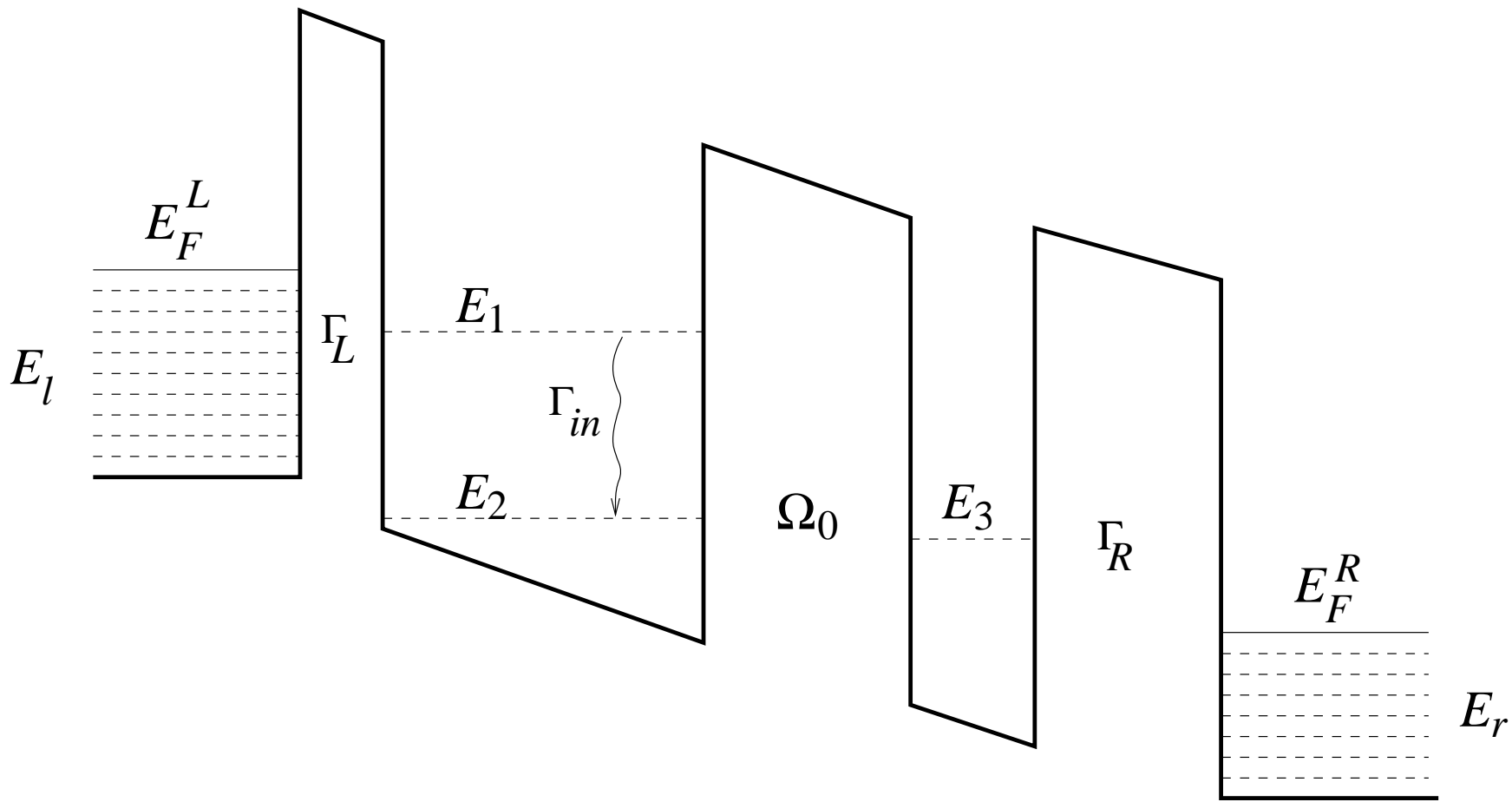
\begin{figure}[h]
\caption{Resonant transport through a double-well structure in the
presence of inelastic process.}
\end{figure}
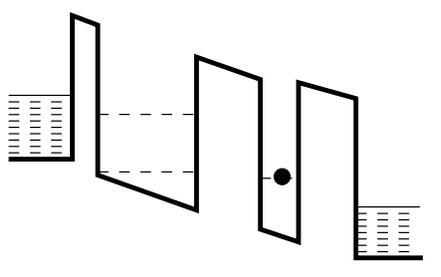
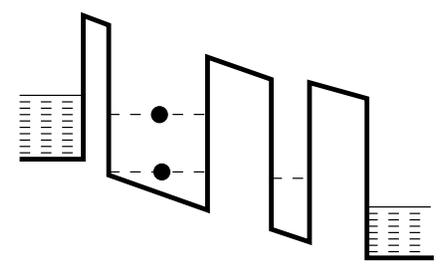
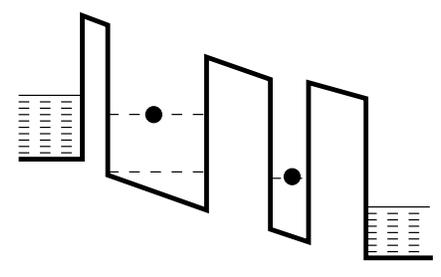
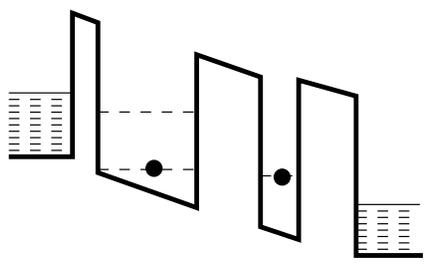
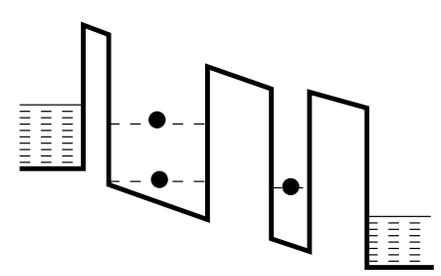
\begin{figure}[h]
\caption{All possible electron states of the device, shown in Fig. 3: 
$|a\rangle$ -- all the levels $E_{1,2,3}$ are empty; $|b\rangle$ -- the upper
level, $E_1$, is occupied; $|c\rangle$ -- the lower level,
$E_2$, is occupied; $|d\rangle$ -- the level $E_3$ is occupied;
$|e\rangle$ -- the levels $E_1$ and $E_2$ are occupied; 
$|f\rangle$ -- the levels $E_1$ and $E_3$ are occupied;
$|g\rangle$ -- the levels $E_2$ and $E_3$ are occupied;  
$|h\rangle$ -- all the levels $E_{1,2,3}$ are occupied.}
\end{figure}
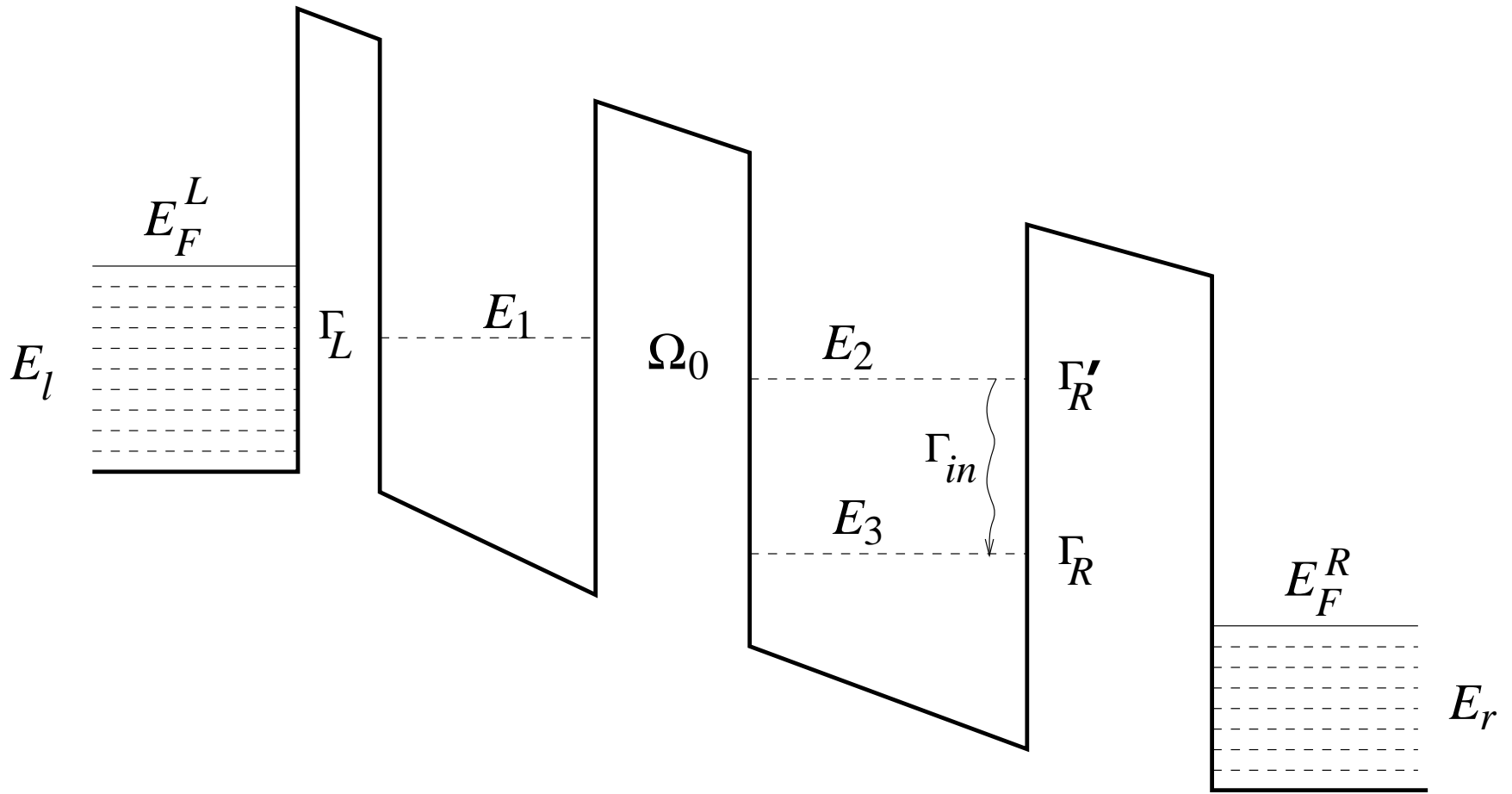
\begin{figure}[h]
\caption{Resonant transport through a double-well structure in the
presence of inelastic process with strong Coulomb blockade effects.}
\end{figure}
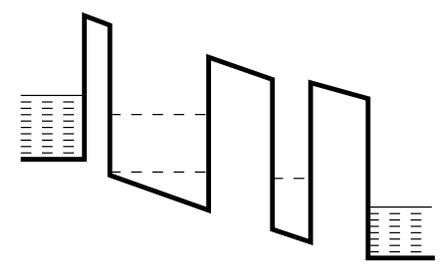
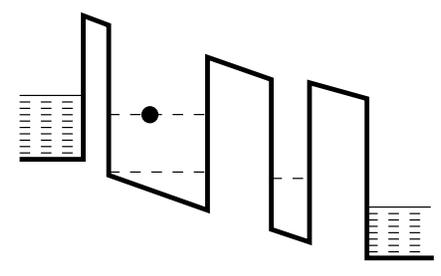
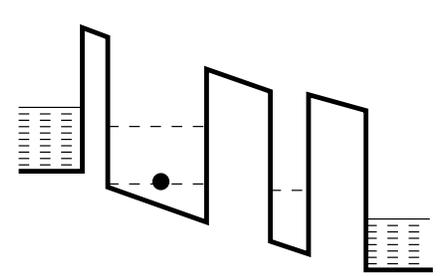
\end{document}